\documentclass[12pt,article,draftcls,letterpaper,onecolumn]{IEEEtran}

 \usepackage{graphicx}
\usepackage{multicol}
\usepackage{amssymb}
\usepackage{epsfig}
\usepackage[cmex10]{amsmath}
\usepackage{slashbox}

\usepackage{setspace}
\ifCLASSINFOpdf
\else
\fi
\begin{document}
%
\title{Lower Bound on the BER of a Decode-and-Forward Relay Network Under Chaos Shift Keying  Communication System}
%
%
%

\author{\IEEEauthorblockN{Georges Kaddoum$^1$ \textit{Member, IEEE}, Fran\c cois Gagnon \textit{Member, IEEE}
}
\IEEEauthorblockA{ LaCIME Laboratory, ETS, Montreal, Canada\\
Email: georges.kaddoum[at]lacime.etsmtl.ca}
}

\author{Georges Kaddoum$^{1}$ \textit{Member, IEEE},  Fran\c cois Gagnon \textit{Member, IEEE}
\thanks{G. Kaddoum and F. Gagnon are with University of Qu\'{e}bec,
Ecole de t\'{e}chnologie supp\'{e}rieure, 1100  Notre-Dame West, Montr\'{e}al  H3C 1K3, Canada (e-mail:
georges.kaddoum@lacime.etsmtl.ca;  francois.gagnon@etsmtl.ca)}

}

\maketitle

\footnotetext [1] {This work has been supported by the Natural Science and Engineering Council of Canada discovery grant $435243-2013$}

\begin{abstract}
This paper carries out the first-ever investigation of the analysis of a cooperative Decode-and-Forward (DF) relay network with Chaos Shift Keying  (CSK) modulation. The performance analysis of DF-CSK in this paper takes into account the dynamical nature of chaotic signal, which is not similar to a conventional binary modulation performance computation methodology. The expression of a lower bound bit error rate (BER) is derived in order to investigate the performance of the cooperative system under independently and identically distributed (i.i.d.) Gaussian  fading wireless environments. The effect of the non-periodic nature of chaotic sequence leading to a non constant bit energy of the considered modulation is also investigated. A computation approach of the BER expression based on the probability density function of the bit energy of the chaotic sequence, channel distribution, and number of relays is presented. Simulation results prove the accuracy of our BER computation methodology.
\end{abstract}



%
\IEEEpeerreviewmaketitle

\section{Introduction}

The growing demand for high data rates has resulted in the need for cooperative communication networks. Cooperative relay networks have emerged as a powerful technique for combatting multipath fading in wireless channels, leading  the performance and the data rate improvements. Consequently, the cooperative network architecture  has received great attention in the  design  and implementation of distributed virtual multiple antenna systems \cite{Lan04}. The transmission between the source and destination is achieved by one or more relays with  a spatial separation needed between  the relays to  ensure spatial diversity.
Two basic cooperative strategies have been proposed in the literature \cite{Tai07} \cite{Shen07}: Amplify-and-Forward (AF) and Decode-and-Forward (DF).

In the AF scenario, the cooperative relay amplifies and forwards the received signal  to the destination.  This scenario is simple where no decoding process is required, but its major drawback is the noise amplification involved \cite{Shen07}. In the DF scheme, the relay decodes the received signal and forwards the decoded message to the destination. We found two strategies in the DF scheme in the literature. The first one is the  Fixed Decode-and-Forward (FDF) strategy, and here, the relay always decodes and forwards the received symbol. In the second strategy named Adaptive Decode-and-Forward relaying (ADF), the relay forwards the received symbol only when it is properly decoded \cite{Tai07}. Therefore, the ADF scheme ensures better bit error rate performances than do the AF and FDF protocols.

For over a decade now, chaotic systems have been attracting a great deal of attention from many researchers, due mainly to the advantages provided by chaotic signals, such as robustness in multipath environments and resistance to jamming \cite{Tam03} \cite{Gal011}. Chaotic signals are non-periodic, broadband, and difficult to predict and to reconstruct. These are properties which coincide with the requirements for signals used in communication systems, particularly for spread-spectrum communication and secure communications \cite{Tam03}  \cite{Wren10} \cite{kadiet}. The possibility of generating an infinite number of uncorrelated chaotic sequences from a given map simplifies the application of these signals in a multi-user case. 

Generally, two different approaches have been adopted for using the chaotic signal in digital communications systems. The first uses the real value of the chaotic signal to modulate the transmitted bits \cite{Tam03}, while the second quantizes the chaotic signal before using it to transmit the data, which offers better performance than conventional spreading systems. Moreover, this approach leads to a loss of chaotic signal properties. Though it is the first approach that is adopted in our paper, the general analysis in the paper can nonetheless be directly adapted for quantized chaotic signals.

Many chaos-based communication systems with coherent and non-coherent receivers are proposed and evaluated; these include chaos shift keying \cite{Tam03}, chaos-based DS-CDMA \cite{Kad09ieee} \cite{kaddoum2013performance},  differential chaos shift keying (DCSK) \cite{Wei11} \cite{kad12cs}, and MC-DCSK \cite{kad13mc}. With coherent receivers, such as CSK system, the chaotic signal is used as a spread sequence to spread the data information signal, while on the receiver side, chaotic synchronization is required to demodulate the transmitted signals \cite{val12}. On the other hand, with differential chaos shift keying (DCSK), chaotic synchronization is not used on the receiver side, where an exact knowledge of the chaotic signal is not required in order to demodulate the transmitted signals. However, the DCSK system is less secure, and does not perform as well as the coherent chaos-based communication systems.

Many papers studied the DF-Relay \cite{6157252},  \cite{965537}. For example, in \cite{965537} they analyze the gain by DF-Relay with general digital modulation scheme theoretically. In this paper, it is considered that digital modulation scheme and the gain by DF-Relay are independent. Since the transmitted bit energy of chaos based modulation schemes is not constant, the BER computation methodology used for conventional digital modulation schemes cannot be applied for Chaos Shift Keying modulation. 

To the best of our knowledge, the DF-CSK system is not proposed or studied in the literature. The aim of this paper is to present a cooperative network under chaos shift keying modulation with DF relay. Once the system is proposed,  the performance of the cooperative system under independently and identically distributed (i.i.d.) Gaussian fading wireless environments is analyzed. In this paper we assume that the cooperative relays properly decode the received message during transmission to obtain the lower bound BER of the scheme. This process is namely, error-free DF at the relay user.

Many assumptions have been made in computing the bit error rate performance of chaos-based communication systems. A widely used approach known as the "Gaussian approximation" considers the sum of dependent variables at the output of the correlator as a Gaussian variable. This approach leads to a low precision of BER approximations, especially when the spreading factor is low \cite{Tam03}. In \cite{Law03}, another exact approach was derived to compute the BER for chaos-based communication systems. Their model takes into consideration the dynamic properties of chaotic sequences, and the performance was derived by integrating the BER expression for a given chaotic map over all possible chaotic sequences for a given spreading factor. This method outperforms the Gaussian approximation, and matches the exact BER. Moreover, as mentioned in \cite{Law03}, this method needs a high computing load. Another accurate simple approach  is used in this paper to compute the exact BER performance of the system \cite{kadiscas12}. This approach proceeds by first estimating the probability density function (pdf) of the bit energy, and then using the computed pdf  to compute the BER expression. 

The originality of this paper is that it proposes for the first time a CSK modulation in a cooperative network, and then extends the BER computation methodology of \cite{Kad09ieee} to be applied to chaos-based cooperative communication.

This paper is organized as follows. In section $2$, we present the CSK communication system. Section $3$ is dedicated to explaining the cooperative DF-CSK network. An analytical performance analysis is presented in section $4$. Simulation results are shown in section $5$, and finally, some concluding remarks are given.

\section{CSK communication system}

In a CSK communication system, the information symbols $(s_{l}= \pm 1)$ with period $T_s$ are spread by a chaotic sequence $x_{k}$. A new chaotic sample (or chip) is generated at every time interval equal to $T_{c}$ ($x_{k}=x(kT_{c})$). 
The emitted signal at the output of the transmitter is:
\begin{equation}\label{e_sig}
u (t) = \sum\limits_{l = 0}^\infty {\sum\limits_{k = 0}^{\beta - 1} {s_l x_{l\beta + k} } } g(t - (l\beta + k)T_c ),
\end{equation}
\\
where the spreading factor $\beta$ is equal to the number of chaotic samples in a symbol duration $(\beta=T_s/T_c)$, and $g(t)$ is the pulse shaping filter. In this paper, we have chosen a rectangular pulse of unit amplitude on $[0, \, T_c] $.
To demodulate the transmitted bits, the received signal is first despread by the local chaotic sequence, and then integrated into a symbol duration $T_s$. Finally, the transmitted bits are estimated by computing the sign of the decision variable at the output of the correlator.

\section{Cooperative system model}

In this section, we present a cooperation strategy for the proposed wireless network. The user in the network can be a source node that sends information to its destination, or a relay node that helps transmit other users' information. We assume that the relay selection of the given user is fixed.  In this case, signal transmission involves two phases, Phase $1$ and Phase $2$. In Phase $1$, each source sends information to its destination, and the information is also received by other users in the network. In Phase $2$, each relay receives the  signal transmitted in Phase $1$. To compute the lower BER bound that can be reached by this network,  we assume that the error-free DF relays always  properly decode the  information received  and forward the decoded symbols to the destination.


\begin{figure}[!htb]
\centering 
\includegraphics[width=7.5cm]{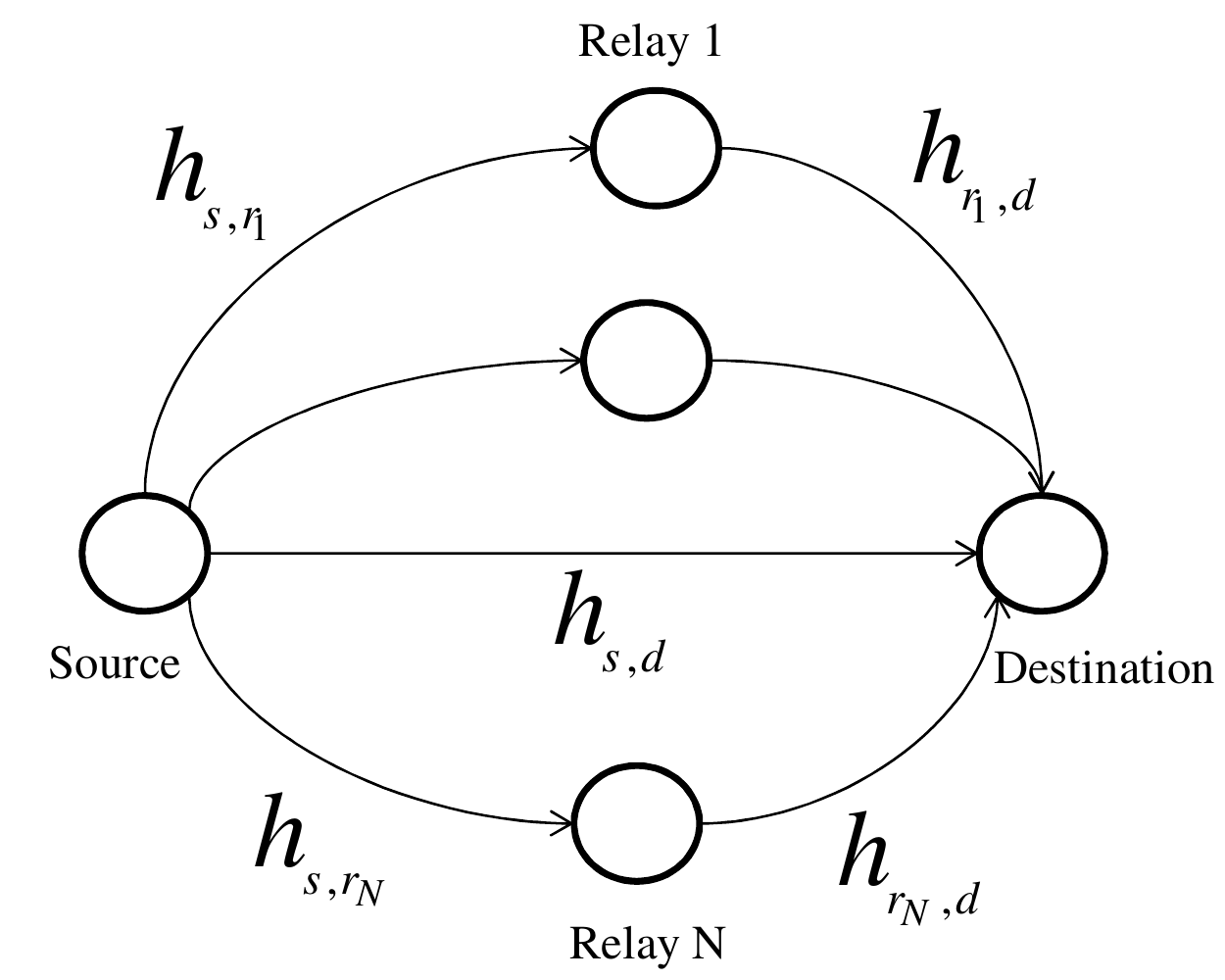}
\caption { System model} 
\label{df_scheme}
\end{figure}

As illustrated in Fig. \ref{df_scheme} each wireless link between any two nodes in the network is subject to fading and additive noise. Although the fading coefficients can be accurately  estimated by the cooperating terminals, the additive noise is unknown to the receiver. For mathematical simplification, we omit in the paper the time  $t$ used in   equation (\ref{e_sig}). In the first phase, the source broadcasts its information to the
destination and the relays. The received signals $y_{s,d}$ and $y_{s,r_{n}}$ at the destination and the $n^{th}$ relay  can respectively be expressed as

\begin{equation} \label{sor1}
y_{s,d}  = \sqrt {P_s } h_{s,d}u  + z_{s,d},
\end{equation}

\begin{equation}\label{sor2}
y_{s,r_n }  = \sqrt {P_s } h_{s,r_n }u  + z_{s,r_n },
\end{equation}
\\
where $u$ is the transmitted spread spectrum information from CSK modulation, $P_s$ is the transmitted power at the source,  $z_{s,d} $ and  $z_{s,r_{n}}$  are the additive white Gaussian noise with variance $\sigma^2$.  In equations (\ref{sor1}) and (\ref{sor2}), $h_{s,d}$ and $h_{s,r_n }$  are the channel coefficients from the source to the destination and to the $n^{th}$ relay, respectively. The channel coefficients are modeled as zero-mean, complex Gaussian random  variables with variances $\sigma_{s,d}^2$ and $\sigma_{s,r_{n}}^2$, respectively. In this paper, we assume that the channel coefficient is maintained constant during the symbol period $T_s$ and vary for each symbol.

Because we assume in our analysis that the relays forward the correct symbols to the destination, the received signal in Phase $2$ at the destination coming from the $n^{th}$ relay can be modeled as

\begin{equation} \label{rey_d}
y_{r_n ,d}  = \sqrt {P_n } h_{r_n ,d}u  + z_{r_{n} ,d},
\end{equation}
\\
where $ h_{r_n ,d}$ is the channel coefficient modeled as a zero-mean, complex Gaussian random  variable with variance $\sigma_{r_{n},d}^2$, $ z_{r_{n} ,d} $ is the additive white Gaussian noise with variance $\sigma^2$, and $P_n$ is the transmitted power at the relay $n$. Note that the  Gaussian model is is considered in many papers studying the performance of DF systems \cite{1683609}, \cite{6506106}. 

In our analysis, we assume that the nodes are spatially separated such that the channels from different links are independent. In addition, the channel coefficients are assumed to be known at the receiver. The destination coherently combines the different received signals from the source and the relays by using the maximum ratio combiner (MRC) \cite{Bren03}.

\section{Performance analysis}

In this section we analyze the lower bound of the BER performance for the proposed model. Knowing the channel state information, the destination detects the transmitted symbols by jointly combining the received signals from the source and relays. The combined signal at the output of MRC can be given by:

\begin{equation} \label{eq_d}
y = a y_{s,d}  + \sum\limits_{j = 1}^N {a_j y_{r_j ,d} },
\end{equation}
\\
where $N$ is the number of relays in the network and $a$ and $a_j$ are determined to maximize the signal-to-noise ratio (SNR) at the output of MRC. Therefore, $a$ and $a_j$ can be given

\[
a = {\sqrt {P_s } h_{s,d}^{*} / \sigma};  \: \: \: \: a_j  = {\sqrt {P_j } h_{r_j ,d}^{*}/ \sigma }
\]
\\
Note that $h_{s,d}^{*} $ is  the complex conjugate of $h_{s,d}$ and  $ h_{r_j ,d}^{*}$ is the complex conjugate of $ h_{r_j ,d}$.

To decode the transmitted symbol $s_l$, the received signal $y$ is first multiplied by an exact replica of the spreading code, and then summed over the bit duration $\beta T_c$. The decision variable for a given symbol $l^{th}$ is:


\begin{equation} \label{des_var}
\begin{array}{l}
Ds_l  = \\
\mathrm{sign}\left( { \left[ {P_s  \vert h_{s,d}\vert^2  + \sum\limits_{j = 1}^N {P_j \vert h_{r_j ,d} \vert^2   } } \right]E_b^{(l)}  + w_{s,d}^{(l)}  + \sum\limits_{j = 1}^N {w_{r_j ,d}^{(l)} } } \right),
  \end{array}
\end{equation}
\\
where $w_{s,d}^{(l)}$  and ${w_{r_i ,d}^{(l)} }$ are the  noises  after despreading and integration and $E_{b}^{(l)} = T_c \sum\limits_{k = 0}^{\beta - 1} {(x_{l\beta +k} )^2 }$ is the transmitted energy of the $l^{th}$ data symbol. 

Because of the non-periodic nature of the chaotic sequence, the energy  $E_{b}^{(l)}$ is not constant for all transmitted bits. Computing the performance requires knowledge of the pdf of the received bit energy multiplied by the channel coefficients. In our case, the performance requires knowledge of the pdf $\alpha$, where $\alpha=\left[ {P_s \vert h_{s,d} \vert^2  + \sum\limits_{j = 1}^N {P_j \vert h_{r_j ,d}\vert^2 } } \right]E_b^{(l)} $. Now, since $\alpha$ which represents the received energy, depends on the particular distribution of the  transmitted energy, $E_b^{(l)}$, and the channels coefficients, the computation of the actual pdf of $\alpha$ leads to a non-standard distribution, which has no analytical form. We therefore  propose that this distribution be approximated by a known distribution which will allow the existing results on the computation of binary modulation schemes to be used to derive a closed-form expression of our system.

\subsection{Chaotic generators}\label{ko_maps}
The  BER computation methodology developed in this article is not restricted to a particular chaotic map, or to a particular type of
chaotic map. We chose the following two families of chaotic maps, widely considered in chaos-based communication systems \cite{Tam03},
the Chebyshev polynomial function of order $2$,  and a piecewise linear map (PWL), defined by:
\begin{enumerate}
\item [1-] CPF map: $x_{k+1}=1-2x_{k}^{2}$
\end{enumerate}

\begin{enumerate}
\item [2-] PWL map: $\left\{ \begin{array}{l}
 z_k  = L\left| {x_k } \right| + \phi \left[ {\bmod 1} \right] \\
 x_{k + 1}  = \textrm{sign}(x_k )(2z_k  - 1) \\
 \end{array} \right.$
\end{enumerate}
The PWL map is parameterized by a positive integer $L$ and a real
number $\phi$ ($0<\phi<1$ ). In this article, these
parameters are set to $L=3$ and $\phi=0.1$.

The  chaotic sequences used here are zero-mean and with a unit variance $\sigma_x^{2}=1$.

\subsection{BER computation methodology}

Our performance computation is not based on Gaussian approximation, where the transmitted bit energy is considered as constant. Note that the Gaussian approximation provides a good estimation of performance just when the spreading factor is high, but for a low spreading factor, this approximation is not valid ~\cite{Kad09ieee}. 

The statistical properties of the decision variable $Ds_l$ are computed for a fixed bit $l^{th}$. The mean of $Ds_l$ is reduced to the term given in equation (\ref{m1}) because the elements of the noise component and the wireless channels are independent and zero-mean:

\begin{equation}\label{m1}
E\left[ {Ds_l } \right] = \left[ {P_s \vert h_{s,d} \vert^2  + \sum\limits_{j = 1}^N {P_i \vert h_{r_j ,d}\vert^2 } } \right]E_b^{(l)} s_l, 
\end{equation}
\\
where $E[.]$ is the mean operator.

The variance of the decision variable is:
\begin{equation}\label{v1}
\small
{\mathop{\rm Var} \left[ {Ds_{l} } \right]}= E\left[ {\left( {Ds_{l} } \right)^2 } \right] - E\left[ {Ds_{l} } \right]^2,  \\ 
\end{equation}
\\
where $\mathrm{Var}[.]$ is the variance operator. 

Since the noise  and the channels coefficients are uncorrelated and independent of the chaotic sequences, the Gaussian noise samples are independent. In addition, we assume that the additive noise variances are the same in all links. The variance of the decision variable is thus the sum of the variances of the two noise expressions $w_{s,d}^{(l)}$, and $  \sum\limits_{j = 1}^N {w_{r_j ,d}^{(l)}} $ in equation (\ref{des_var}).

The variance of the first noise  component is:

\begin{equation} \label{var1}
\mathrm{Var}  \left[ w_{s,d}^{(l)}\right]= \mathrm{Var} \left[ {\sum\limits_{k = 0}^{\beta  - 1} {\sqrt {P_s } z_{s,d}^{(l\beta  + k)} h_{s,d} x_{l\beta  + k} } } \right] \cdot
\end{equation}

Since the chaotic and the noise signals are zero-mean, the variance of $w_{s,d}^{(l)}$ is reduced to:
\begin{equation} \label{var01}
\mathrm{Var}  \left[ w_{s,d}^{(l)}\right] =E\left[ {\sum\limits_{k = 0}^{\beta  - 1} {\sqrt {P_s } z_{s,d}^{(l\beta  + k)} h_{s,d} x_{l\beta  + k} } } \right]^2 \cdot
\end{equation}

By considering the uncorrelated and independent hypothesis of the channel coefficients, the chaotic and noise samples, the variance can be written:

 \begin{equation} \label{var01}
\mathrm{Var} \left[ w_{s,d}^{(l)}\right] = \beta P_s \frac{{N_0 }}{2} \vert h_{s,d} \vert^2 E\left[  x_{l\beta  + k}  \right]^2,
\end{equation}
\\
where $E_b^{(l)}= \beta E\left[  x_{l\beta  + k}  \right]^2$ is the transmitted bit energy. Then the variance expression of $w_{s,d}^{(l)}$ is:

\begin{equation} \label{var1}
\mathrm{Var} \left[ w_{s,d}^{(l)}\right] = P_s \frac{{N_0 }}{2} \vert h_{s,d} \vert^2 E_b^{(l)} \cdot
\end{equation}

%

The variance of the second noise component $  \sum\limits_{j = 1}^N {w_{r_j ,d}^{(l)}} $ is:

\begin{equation}
\mathrm{Var} \left[ \sum\limits_{j = 1}^N {w_{r_j ,d}^{(l)}} \right] = \mathrm{Var} \left[ {\sum\limits_{k = 0}^{\beta  - 1} {\sum\limits_{j = 1}^N {\sqrt {P_j } z_{r_j ,d}^{(l\beta  + k)} h_{r_j ,d} x_{l\beta  + k} } } } \right] \cdot
\end{equation}
 
In our paper, the channel coefficients are assumed independent from each other and from the noise and chaotic signals. In this case, the variance of  $  \sum\limits_{j = 1}^N {w_{r_j ,d}^{(l)}} $ can be written:

\begin{equation}
\mathrm{Var} \left[ \sum\limits_{j = 1}^N {w_{r_j ,d}^{(l)}} \right] = E\left[ {\sum\limits_{k = 0}^{\beta  - 1} {\sum\limits_{j = 1}^N {\sqrt {P_j } z_{r_j ,d}^{(l\beta  + k)} h_{r_j ,d} x_{l\beta  + k} } } } \right]^2 \cdot
\end{equation}

By analogy, the variance of the second noise component is: 
\begin{equation}
\mathrm{Var} \left[ \sum\limits_{j = 1}^N {w_{r_j ,d}^{(l)}} \right] = \frac{{N_0 }}{2}E_b^{(l)} \sum\limits_{j = 1}^N {P_j } \vert h_{r_j ,d} \vert^2 \cdot
\end{equation}

Finally, the variance of the decision variable is:

\begin{equation}
\mathrm{Var}\left[ {D_s^{(l)} } \right] = \frac{{N_0 }}{2}E_b^{(l)} \left( {P_s \vert h_{s,d} \vert^2  + \sum\limits_{j = 1}^N {P_j } \vert h_{r_j ,d} \vert^2 } \right) \cdot
\end{equation}

\subsection{Computation methodology}

Since the symbols are equally distributed on the set $\{-1,+1\}$, we can deduce from \cite{Sim05} that the BER for symbol $l$, for a given constant bit  energy and channel coefficients is
expressed as:
\begin{equation}
\textrm{BER}_l   =  Q\left( {\frac{{E\left[ {Ds_l } \right]}}{{\sqrt {\mathrm{Var} \left[ {Ds_l } \right]} }}} \right), \\ 
\end{equation}
\\
where the Gaussian $Q$-function is defined as:  $Q(x) = \int\limits_x^\infty  {\frac{1}{{\sqrt {2\pi } }}} e^{\left( {\frac{{-u^2 }}{2}} \right)} du$. Finally , the conditional BER expression is
\begin{equation}
 \textrm{BER}_l = Q\left( {\sqrt {\frac{{2\left( {P_s \vert h_{s,d} \vert ^2 \sum\limits_{j = 1}^N { P_j \vert h_{r_j ,d} \vert^2 } } \right)E_b^{(l)} }}{{N_0 }}} } \right) \cdot \\ \end{equation}

\begin{equation} \label{prob_err}
 \textrm{BER}_l= Q\left( {\sqrt {\frac{{2\alpha }}{{N_0 }}} } \right) \cdot
\end{equation}

The mean BER of the system is obtained by integrating
(\ref{prob_err}) over all possible values of $\alpha$, 
the bit energy and

\begin{equation} \label{eq_2}
\overline{ \textrm{BER}_l }  = \int_0^\infty   Q\left(
\sqrt {\frac{2 \alpha } {N_0}} \right)
p(\alpha) d \alpha \cdot
\end{equation}

Basically, the computation of $\textrm{BER}_l$ should require 
knowledge of the pdf's $p(\alpha)$.  Now, considering the expression (\ref{eq_2}) of
$\overline{\textrm{BER}}_l$, one can note that it has the same form
as the expression of the mean BER obtained in the framework of
mobile radio channels. Indeed, for a BPSK transmission on a radio
channel with gain $\rho$ and with bit energy-to-noise ratio
$E_b/N_0$, the mean BER is expressed as
\begin{equation}\label{ber_radio}
\textrm{BER}_{\textrm{Radio channel}}  = \int_0^\infty {Q\left(
{\sqrt {\frac{{2\rho ^2  E_b}}{{N_0 }}} } \right)} p\left(
\rho \right)d\rho \cdot
\end{equation}
For Rayleigh, Nakagami, gamma  or Rician
channels, there exist closed-form expressions for
(\ref{ber_radio}) \cite{Sim05}. Since expression (\ref{eq_2}) is similar to
expression (\ref{ber_radio}), these previous results on
(\ref{ber_radio}) can be used to obtain an analytical form of
the integral (\ref{eq_2}). This article is partly intended to derive an analytical expression of $\overline{\textrm{BER}}_l$.

The distribution approximation has been investigated by plotting
the histogram of the variable $\alpha$, and by comparing it with the
theoretical pdf's of the Rayleigh, Rician, and Nakagami, and gamma distributions,
respectively. The parameters have been defined such
that the moments of the theoretical pdf match the estimated moments
of $\alpha$. Note that other approximation strategies could be
considered. For instance, one could compute the parameters which
minimize the distance between the theoretical pdf and the histogram
in the least-square (or in any other) sense. However, in such
cases, the parameters could only be obtained numerically. However, the method chosen in this paper allows us to obtain an
analytical expression of the pdf parameters as a function of the
system parameters. To that end, the  generalized  gamma  distribution function  given in \cite{Al05} \cite{Mal09} is considered. Based on the estimated values of its parameters, the distribution of the histogram can be approximated. The  generalized  gamma  distribution function is:

\begin{equation} 
p_\alpha  (\alpha ) = \frac{2v}{{(\Omega/m )^m \Gamma (m)}}\alpha ^{2mv - 1} \exp \left( { - \frac{m\alpha^2v }{\Omega }} \right),\,\,\,\,\,\,\alpha  \ge 0
\end{equation}
\\
where $v$ is the shape parameter, $\Omega$ is the scaling parameter, $m$ is the fading parameter, and $\Gamma (.)$ is the Gamma function. 

By changing the two parameters $v$ and $m$, different fading models can be obtained. The Rayleigh ($m = v = 1$), Nakagami-m ($v = 1$), Weibull ($m = 1$), and lognormal ($m \to \infty $ , $ v \to  0$) distributions are special or limiting cases of the generalized gamma distribution. A full description of the parameter estimations ($v$, $m$ and  $\Omega$ ) is given in  \cite{Rei11}.

\begin{figure}[!htb]
\centering 
\includegraphics[width=8cm]{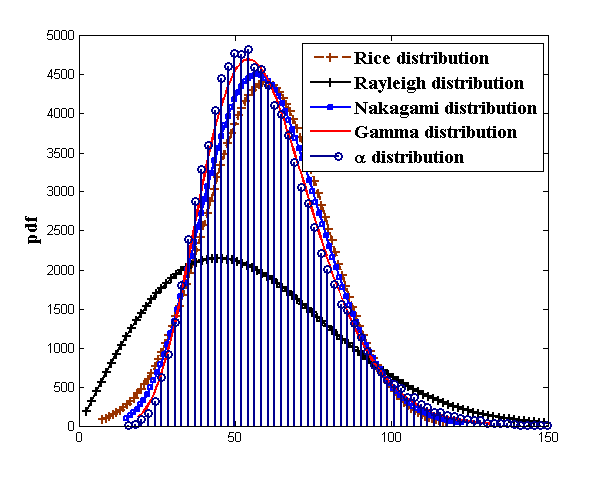}
\caption {Histograms of variable $\alpha$ for CPF chaotic map, six independent channels, spreading factor equal to $10$, and the corresponding approximated pdf's distributions.} 
\label{Hist_cpf}
\end{figure}

\begin{figure}[!htb]
\centering 
\includegraphics[width=8.2 cm]{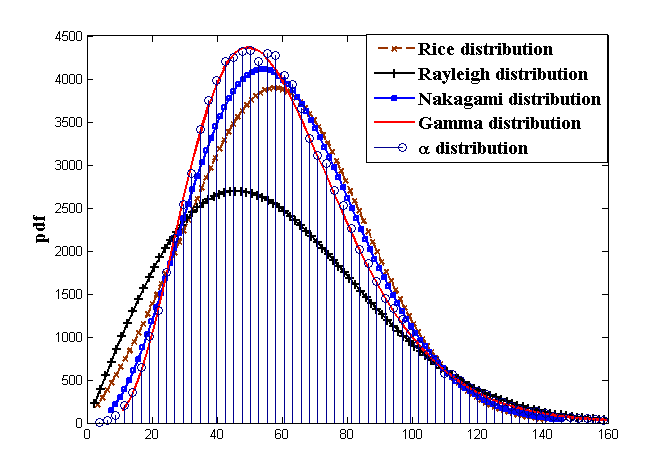}
\caption {Histograms of variable $\alpha$ for PWL chaotic map, six independent channels, spreading factor equal to $10$, and the corresponding approximated pdf's distributions.} 
\label{Hig_pwl}
\end{figure}

Many different sets of system parameters (including different chaotic sequences) have been tested for this
approximation, for different numbers of relays. In each case, the gamma distribution has always best
approximated the histogram of $\alpha$. This observation is illustrated
in Figs. \ref{Hist_cpf} and  \ref{Hig_pwl}, which show the histogram of $\alpha$ of the CPF and  PWL map for given
system parameters, along with the theoretical Rayleigh, Rician, gamma, and Nakagami  pdfs. The histogram of each sequence is obtained for $10^{6}$ bits with  spreading factor equal to $\beta=10$, and for six different fading Gaussian channels (5 relays) with unit variance.


The performance of our communication system, in terms of bit error rate (BER) depends on the statistics of the SNR. The instantaneous SNR per received symbol is $\gamma=  \frac{\alpha }{N_0}$ and the average SNR is $ \bar \gamma  = \frac{{E\left[ \alpha  \right]}}{{N_0 }}$. As computed in  \cite{Sim05}, the BER for binary phase shift keying with coherent detection  is:

\begin{equation}
\textrm{BER}= \frac{1}{\pi }\int_0^{\pi /2} {\exp \left( { - \frac{\gamma }{{\sin \left( \theta  \right)^2 }}} \right)d\theta } \cdot
\end{equation}

The average BER  for gamma distribution is computed in  \cite{Mal09} where the performance expression is derived with the aid of moment generating function (MGF) approach.

\begin{equation} \label {ber_f}
\overline {\textrm{BER}}  = \frac{1}{\pi }\int_0^{\pi /2} {M_\gamma  \left( { - \frac{1}{{\sin \left( \theta  \right)^2 }}} \right)d\theta } \cdot 
\end{equation}
\\
The MGF $M_\gamma$ is computed in \cite{Mal09}.

\begin{equation} \label{mgf}
M_\gamma  \left( \varepsilon  \right) = \sum\limits_{n = 0}^\infty  {\frac{{( - 1)^n }}{{n!}}\bar \gamma ^n \frac{{\Gamma (m + n/v)\Gamma ^{n - 1} (m)}}{{\Gamma ^n (m + 1/v)}}} \varepsilon ^n,
\end{equation}
\\
where $m$ and $v$ are the parameters estimated from the generalized gamma distribution given in \cite{Rei11}. 

The infinite series of the MGF given in equation (\ref{mgf}) is not guaranteed to converge for all values of $\varepsilon $. Many approximations can solve this problem like Pad\'{e} approximation technique to obtain efficiently the limiting behavior of a power series in compact rational function form. The convergence point is not the aim of our paper, however this problem is  deeply discussed in \cite{Amin94} and \cite{776592}.

\section{Simulation}

\begin{figure}[!htb]
\centering 
\includegraphics[width=8cm]{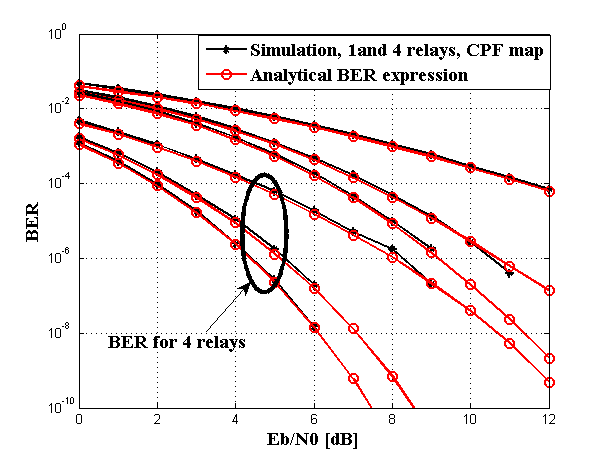}
\caption {Average BER of one  and $4$ DF relays with CPF map, and different spreading factors ($\beta= 5, \: 15, \: 30$) } 
\label{cpf_1_r_5_15_30}
\end{figure}

\begin{figure}[!htb]
\centering 
\includegraphics[width=8 cm]{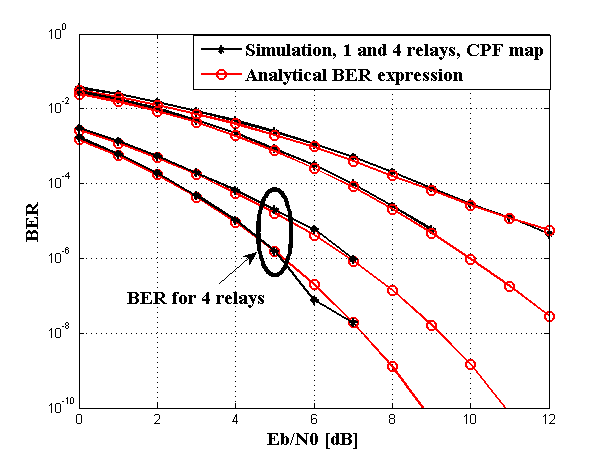}
\caption {Average BER of one and $4$  DF relays with PWL map, and different spreading factors ($\beta=: 15, \: 30$)} 
\label{pwl_1_r_5_15_30}
\end{figure}

We performed computer simulations of the decode-and-forward cooperation system  using a CSK modulation to illustrate the above theoretical analysis. In our simulations, the average BER performances are plotted for different spreading factors ($\beta= 5, \: 15, \: 30$), and normalized power ($P_s=P_j=1$). Figs. \ref{cpf_1_r_5_15_30} and  \ref{pwl_1_r_5_15_30} compared the lower bound BER formulation of equation (\ref{ber_f}) solved numerically and the simulated performance in the case of CPF and PWL maps. Clearly, the theoretical results matched the simulation results for all spreading factors and numbers of relays. This good match therefore validates approximation of the pdf and the computation methodology.

%

\section{Conclusion}

In this paper, a chaos shift keying technique in a cooperative network is proposed and evaluated under error-free DF at the relay user assumption. A chaotic sequence is used in the communication system thanks to the increased security level of the transmission it brings, at low cost. Once the system is proposed, a performance analysis is performed under a fading Gaussian channels. The lower bound BER derivation takes into account the non-periodic nature of chaotic sequences, the channel model, and the number of relays. The BER computation method is based on an approximation of the probability density function of the bit energy multiplied by the sum of different channel attenuations. This approximation is obtained using the generalized gamma distribution, whose appropriate parameters are derived in function of the variable $\alpha$. There is an excellent match between the analytical and the simulated BERs. However, since the general gamma distribution function is used to estimate the pdf, other chaotic maps  with other types of distributions could be considered, using the same principle, in order to derive an analytical BER expression.









\bibliographystyle{IEEEtran}
\bibliography{bibliographie}

\end{document}